\documentclass[reviewcopy]{elsarticle}
\usepackage[reviewcopy]{adndt}
\usepackage{longtable}

\usepackage{amsmath}

\usepackage{amssymb}
\usepackage{graphicx}
\biboptions{square,sort&compress}
\bibpunct[]{[}{]}{,}{n}{}{;}
\begin{document}

\begin{frontmatter}

\journal{Atomic Data and Nuclear Data Tables}

%% Author, fill in article title here

\title{Multi-configuration Dirac-Hartree-Fock calculations of forbidden transitions within the $3d^k$ ground configurations of highly charged ions ($Z=72-83$)}

%% Fill in author list here
\author[hb]{Z.L. Zhao}
\author[hb,IAPCM,IMP]{K. Wang\corref{corl}}
\author[IAPCM]{S. Li}
\author[IMP]{R. Si}
\author[IMP]{C.Y. Chen}
\author[nudt]{Z.B. Chen}
\author[IAPCM]{J. Yan\corref{corl}}
\author[NIST]{Yu. Ralchenko\corref{corl}}
\cortext[corl]{Corresponding Author} \ead{wang$_{-}$kai10@fudan.edu.cn (K.Wang), yan$_{-}$jun@iapcm.ac.cn (J.Yan),  yuri.ralchenko@nist.gov}
\address[hb]{Hebei Key Lab of Optic-electronic Information and Materials, The College of Physics Science and Technology, Hebei University, Baoding 071002, China}
\address[IAPCM]{Institute of Applied Physics and Computational Mathematics, Beijing 100088, China}
\address[IMP]{Shanghai EBIT Lab, Institute of Modern Physics, Department of Nuclear Science and Technology, Fudan University, Shanghai 200433, China}

%\address[ch]{Department of Radiotherapy, Shanghai Changhai Hospital, Second Military Medical University, Shanghai 200433, People?s Republic of China}
%\address[PK]{Center for Applied Physics and Technology, Peking University, Beijing 100871, China}
%\address[jt]{Collaborative Innovation Center of IFSA (CICIFSA), Shanghai Jiao Tong University, Shanghai 200240, China}
\address[nudt]{College of Science,  National University of Defense Technology, Changsha 410073, China}
\address[NIST]{National Institute of Standards and Technology, Gaithersburg, Maryland 20899-8422, USA}
%\date{31.07.2016}

\begin{abstract}
	Extensive self-consistent multi-configuration Dirac-Hartree-Fock (MCDHF) calculations are performed for the $3s^2 3p^6 3d^k$ ($k=1-9$) ground configurations of highly charged ions ($Z=72-83$). Complete and consistent data sets of excitation energies, wavelengths, line strengths, oscillator strengths, and magnetic dipole (M1) and electric quadrupole (E2) transition rates among all these levels are given. We have compared our results with the results available in the literature and the accuracy of the data is assessed. We predict new energy levels and transition probabilities where no other experimental or theoretical results are available, which will form the basis for future experimental work.
\end{abstract}

\end{frontmatter}

%%% Keywords and subject classification are not used in ADNDT
%%%\begin{keywords}
%%%Insert list of keywords here.
%%%\end{keywords}

%%% The table of contents should start a new page. This command will
%%% automatically pull all the unstarred \section, \subsection and
%%% \subsubsection titles into the Contents. Starred versions need to be
%%% done manually using
%%%            \addcontentsline{toc}{[[sub]sub]section}{Section title}
%%% at the correct place. Examples are given below.

%%% The lists of data figures and data tables are created automatically
%%% by the \listofDfigures and \listofDtables commands.

\newpage

\tableofcontents \listofDtables \listofDfigures \vskip4pc

%%%% Authors begin text of article here %%%
\section{Introduction}\label{sec:intro}
Highly charged ions (HCI) of high-$Z$ elements are currently the subject of extensive research due to their great importance in atomic physics and fusion applications. From a theoretical viewpoint, complex spectra of high-$Z$ elements provide a crucial test for advanced theories of atomic structure in a regime where strong relativistic and quantum-electrodynamic effects greatly affect ion properties\cite{Ralchenko.2011.V83.p32517,Osin.2012.V66.p286,Ralchenko.2013.VT156.p14082}. As for fusion applications, heavy elements %such as tantalum ($Z=73$), tungsten ($Z=74$) and gold ($Z$=79)
are routinely used in fusion environments\cite{Fischer.2009.V386.p789,Hawryluk.2009.V49.p65012,Cook.2008.V26.p479}, and their spectra provide
important diagnostic information on plasma parameters. For instance, tungsten ($Z=74$) is being implemented as the plasma-facing material in the ITER divertor region. As a result, tungsten ions will be transported from the relatively cold divertor region to the plasma core with temperatures on the order of 20 keV. Tungsten atoms will ionize to M-shell and L-shell charge states, and hence useful information for plasma diagnostics can be derived from spectral lines of these tungsten ions~\cite{Clementson.2014.V100.p577}. Also, alloys containing tantalum (Z = 73) are considered to be another potential candidate for a plasma-facing material in the ITER, and gold (Z = 79) hohlraums are often employed as targets for inertial confinement fusion studies. Therefore, physical properties of Ta and Au under the influence of hot plasmas are also being examined~\cite{Fischer.2009.V386.p789,Hirai.2003.VT103.p59,Cook.2008.V26.p479,Glenzer.2012.V54.p45013}. 

One of the most notable characteristics of HCI emission is that forbidden transitions, which are barely seen in light elements, play an important role in plasma diagnostics because the corresponding radiation intensities are often very sensitive to electron temperature and density. An effort of investigation of extreme-ultraviolet (EUV) spectra from HCI of heavy elements has been initiated at the National Institute of Standards and Technology (NIST). Various forbidden lines are often the most prominent features in such spectra\cite{Utter.2000.V61.p30503,Ralchenko.2006.V74.p42514,Ralchenko.2008.V41.p21003,Ralchenko.2011.V83.p32517,Osin.2012.V66.p286,Ralchenko.2013.VT156.p14082}. This paper presents progress in ab initio atomic data and theoretical spectra of forbidden transitions within the $(1s^2 2s^2 2p^2) 3s^2 3p^6 3d^k$ ($k=1-9$) ground configurations of highly charged ions ($Z=72-83$). The structures are calculated using the multiconfiguration Dirac-Hartree-Fock (MCDHF) and subsequent relativistic configuration interaction (RCI) approach. %implemented within a new release~\cite{Jonsson.2013.V184.p2197} of the GRASP2K code~\cite{Jonsson.2007.V177.p597}.
The calculations yield energy levels, %transition energies,
wavelengths, radiative rates, weighted oscillator strengths, and line strengths.

There exist many experimental investigations of forbidden transitions within the $3s^2 3p^6 3d^k$ ($k=1-9$) ground configurations in these M-shell ions with measurements from EBITs at NIST~\cite{Porto.2000.V61.p54501,Ralchenko.2007.V40.p3861,Ralchenko.2008.V41.p21003,Ralchenko.2011.V83.p32517,Draganic.2011.V44.p25001,Osin.2012.V66.p286,Ralchenko.2013.VT156.p14082} and at the Lawrence Livermore National Laboratory~\cite{Utter.2000.V61.p30503}. There are also several theoretical studies (see e.g., Refs. \cite{Fournier.1998.V68.p1,Quinet.2011.V44.p195007,Ralchenko.2011.V83.p32517,Osin.2012.V66.p286,Clementson.2014.V100.p577,Guo.2015.V48.p144020}) on forbidden lines within the ground configurations of these highly charged ions with an open 3d shell. The most comprehensive set of data is presented by Ralchenko $et~al.$~\cite{Ralchenko.2011.V83.p32517,Osin.2012.V66.p286} and~\citet{Guo.2015.V48.p144020}, who modeled the spectra of Hf ($Z=72$), Ta ($Z=73$), W ($Z=74$), and Au ($Z=79$) using the Flexible Atomic Code (FAC)~\cite{Gu.2008.V86.p675}. The present work extends the database for forbidden lines in highly charged ions relevant to fusion plasma diagnostics.  Extensive comparisons with available theoretical and experimental data are discussed.

\section{Calculations}\label{sect:cal}
Energy levels and transition properties for the forbidden transitions within the $3s^2 3p^6 3d^k$ ($k=1-9$) ground configurations of highly charged ions ($Z=72-83$) were obtained using the MCDHF and subsequent RCI approach (MCDHF/RCI) implemented within a new release~\cite{Jonsson.2013.V184.p2197} of the GRASP2K code~\cite{Jonsson.2007.V177.p597}. In this method, the atom is represented by atomic state functions (ASF) $\Psi(\gamma J )$, which are linear combinations of a number of configuration state functions(CSF):
\begin{equation}\label{MCDF01}
\centering
\Psi(\gamma J )= \sum_{i} c_{i}\Phi(\gamma_{i} J ),
\end{equation}
where the CSFs $\Phi(\gamma_{i} J ) $ are constructed from single-electron Dirac orbitals, which are based on the well known rules of symmetry (parity and angular momenta)~\cite{Grant.1980.V21.p207}. The CSFs are generated as all possible combinations of orbitals in an active set (AS), according to certain constraints, to form a restricted active space (RAS) of CSFs. The first step is to set the reference configurations, which are the $ 3s^{2}3p^{6}3d^{k} $ (with $ k=1,\ldots,9 $) configurations in our case, in the future labeled Dirac-Fock (DF). Then the multi-configuration expansions are obtained through single and double excitations from the $ 3s $, $ 3p $ and $ 3d $ shells of the reference configurations up to a $ n_{max}l_{max} $ orbital with $l_{max}=min(n-1,5)$. Because of the rapid expansion of the configuration space, especially for the Ti-, V- and Cr-like ions, we adopt different expansion techniques for different ionic stages: i) for K- and Co-like ions, $ 3l^{ 8+k } $ $ + $ $ 3l^{8+k-1} 4l $ $ + $ $ 3l^{ 8+k-2} 4l^{2} $ $ + $ $3s^{2} (3p,3d)^{6+k-1} nl $ $ + $ $3s^{2} (3p,3d)^{6+k-2} nl ^{2}$ with $n=5,6,7$, and $ k=1 $ or $9$ (here $ nl^{m}$ means that $m$ electrons are arbitarily filled in the $n$ complex ); ii) for Ca-, Sc-, Mn- and Fe-like ions, $ 3l^{8+k } $ $ + $ $ 3l^{8+k-1} 4l $ $ + $ $ 3l^{8+k-2} 4l^{2} $ $ + $ $3s^{2} (3p,3d)^{6+k-1} 5l $ $ + $ $3s^{2} (3p,3d)^{6+k-2} 5l ^{2}$ $ + $ $3s^{2}3p^{6}3d^{k-1} nl $ $ + $ $3s^{2}3p^{6} 3d^{k-2} nl ^{2}$ with $n=6,7$, and $ k=2, 3, 7 $ or $8$; iii) for Ti-, V- and Cr-like ions, configuration expansions are similar to those of Ca-like ion, but with the highest $ n $ up to 6. In order to further reduce the size of the configuration interaction (CI) matrix, we only consider those of the configurations in the above space that directly interact with the reference configurations. Those configurations are used to optimize the atomic orbital (AO) wave functions  via the multi-configuration self-consistent-field (MCSCF) method. The AO wave functions with the principle quantum numbers $ n=1, 2, 3$ are optimized together as spectroscopic orbitals with $ n-l-1 $ nodes by minimizing the weighted sum (with statistical weights) $\mathcal{F}$ of all energy levels of ground configurations $3s^{2}3p^{6}3d^{k}$ together with the $ 3l^{8+k } $ configurations. Then the AOs of $n=1, 2, 3$ are fixed, and AOs for $n=4$ are optimized by minimizing the same $\mathcal{F}$ within the above configuration space expanded to $ n_{max} =4 $. In succession, with the AOs of $ n \leq 4$ fixed, the AOs are extended to $n_{max} = 5,6,7$ from $n$ to $n + 1$, and the AOs  with $n \geq 4$ are optimized as pseudo-orbitals. %in order to take into account the short-range dynamic correlation better with the limited CI basis set.

Table~\ref{table1} displays the computed excitation energies as functions of the increasing active sets and multireference sets for K-like W$^{55+}$ , Ti-like W$^{52+}$, and Fe-like W$^{48+}$ ions. The first column in the table represents the largest principal quantum number of the active set involved in each step of the MCDHF calculations. As can be seen from Table~\ref{table1}, the MCDHF excitation energies have converged in the last step ($n=7$ for W$^{48+}$ and W$^{55+}$ ions, and $n=6$ for W$^{52+}$ ion). For instance, the contributions to the energy of the $(d_{+}^{1})_{5/2}$ level, due to electron correlation effects with the $n=3, 4, 5, 6, 7$ configurations, are 1176, -318, -9, 17, 7 cm$^{-1}$, respectively. The total contribution 810 cm$^{-1}$, relative to the DF energy, is about 0.13~\%. The computed results have also converged for the other ions in the present case when the configurations listed in the previous paragraph were included in the MCDHF calculations, which is not shown in detail here. Furthermore, the relative share of correlation effects in the total excitation energies becomes smaller with increasing Z.

After the AOs are obtained, the same configuration spaces for K- and Co-like ions are used to construct the RCI matrix to calculate the energy levels of ground configurations $ 3s^{2}3p^{6}3d^{k} $ and their eigenvectors with consideration of the Breit interaction as well as the dominant quantum-electrodynamical (QED) corrections - vacuum polarization, finite nuclear mass corrections and self-energy corrections. Larger configuration spaces with additional single and double excitations from the $3p$ shell to the $n=6,7$ orbitals for Ca-, Sc-, Mn- and Fe-like ions, and to the $n=6$ orbitals for Ti-, V- and Cr-like ions, are used in RCI calculations to take into account more core-excitation correlations, which are important for convergence and accuracy of calculated energy levels and transition properties.

In Table~\ref{table1} we list the energy levels of K-like W$^{55+}$ , Ti-like W$^{52+}$, and Fe-like W$^{48+}$ ions, obtained from the MCDHF/RCI calculations with the inclusion of Breit and QED effects. For a majority of levels, our Breit and QED corrected excitation energies are lower than the corresponding Coulomb energies by about 2~\%. Furthermore, inclusion of the Breit and QED corrections has resulted in a slightly different level orderings, such as for levels 28/29 of Ti-like W$^{52+}$ ion. Elements with nuclear charge $Z = 72-83$ are heavy, and hence relativistic effects should be comparatively more important than the electron correlation effects.

Furthermore, to assess the accuracy of the MCDHF/RCI results we have performed further calculations using a combined RCI and many-body perturbation theory (MBPT) approach implemented in the FAC code~\cite{Gu.2005.V89.p267,Gu.2006.V641.p1227,Gu.2008.V86.p675}. In the MBPT approach, the Hamiltonian is taken to be the no-pair Dirac-Coulomb-Breit Hamiltonian $H_{DCB}$. The key feature of the RCI and MBPT approach is the partition of the Hilbert space of the system into two subspaces, i.e., a model space \emph{M} and an orthogonal space \emph{N}. The eigenvalues of $H_{DCB}$ can be obtained through solving the eigenvalue problem of a non-Hermitian effective Hamiltonian in the model space \emph{M}. Taking the first-order perturbation expansion of the effective Hamiltonian within the Rayleigh-Schr{\"o}dinger scheme, it consists of two parts: one is the exact $H_{DCB}$ matrix in the model space \emph{M}, and the other includes perturbations from the configurations in the \emph{N} space up to the second order for the level energies of interest. In the present calculation, the model space \emph{M} contains  configurations $3s^{2}3p^{6}3d^{k}$ with $k=1,\ldots,9$ from K-like to Co-like ions. The \emph{N} space contains all configurations formed by single and double virtual excitations of the \emph{M} space. For single excitation, configurations with $n\leq 150$ and $l\leq \min{(n-1,25)}$ are included. For double excitations, configurations with the inner electron promotion up to $n = 65$ and the outer electron up to $n^\prime = 200$ are considered.

We start the energy structure calculations using an optimized local central potential, which is derived from a Dirac-Fock-Slater self-consistent field calculation with the configurations from the \emph{M} space. We then perform the MBPT calculations to obtain level energies and radiative transition properties, such as transition wavelengths, line strengths, oscillator strengths, and radiative rates of all transitions among the states %in the \emph{M} space
by using the length form. In addition to the Hamiltonian $H_{DCB}$, several high-order corrections such as the finite nuclear size, nuclear recoil, vacuum polarization, and electron self-energy corrections, are also taken into account in the calculations.

\section{Results and discussion}\label{sect:cp}
%\section{Comparisons}
\subsection{Energy levels and wavelengths}
The energy levels obtained using the MCDHF/RCI method described above for the $3s^2 3p^6 3d^k$ ($k=1-9$) ground configurations of highly charged ions ($Z=72-83$) are presented in Table~\ref{table2}.  The symmetry $J^{\pi}$ of the level, i.e.  the total angular momentum and parity, is  given in Table~\ref{table2}. The leading component  of the eigenvector in the \emph{jj}-coupling scheme is used as a label for the energy levels given in Table~\ref{table2}. In the notations adopted, $d_{-}$ corresponds to the principal quantum number $n=3$ and  $j_{-} = 3/2$, while $d_{+}$ is written for $n=3$ and $j_{+} = 5/2$. The subscript after the right parenthesis is the total $J$ value of the term within the parentheses. The notation $3s^2 3p^6$ is omitted.

In terms of experimental work and accurate calculations, W ions are currently the most studied ones among all the ions considered here. Therefore, the results from the present calculations are first validated against experimental data and other theoretical predictions for W. In Table~\ref{table1} excitation energies for W$^{48+}$, W$^{52+}$, and W$^{55+}$ ions from the MCDHF/RCI and MBPT calculations in this work are compared with calculated excitation energies from the previous calculations (MCDHF/RCI2) by ~\citet{Quinet.2011.V44.p195007}, and from the calculations by Ralchenko $et~al.$~\cite{Ralchenko.2011.V83.p32517}, using the RCI theory implemented in the FAC code. Collected in Table~\ref{table1} are also observed experimental excitation energies from Ralchenko $et~al.$~\cite{Ralchenko.2011.V83.p32517}, which have been included in the NIST database~\citep{Kramida.2014.V.p}. The MBPT values from~\citet{Guo.2015.V48.p144020} have accuracy similar to the present MBPT results and are therefore omitted in Table~\ref{table1}. The experimental values are available for 11 levels.  Our MCDHF/RCI  calculations agree with experimental values better than 0.2~\% for 9 levels. The largest difference is 0.4~\% for the level $((d_{-}^{3})_{3/2},(d_{+}^{5})_{5/2})_{2}$ of W$^{48+}$. The mean relative energy difference between calculations and observations are 0.17~\% for MCDHF/RCI in this work, 0.16~\% for MBPT, and 0.30~\% for MCDHF/RCI2. Because of the larger extent of electron correlation effects considered in the present MCDHF/RCI calculations
via including larger multiconfiguration expansions, our results are slightly better than those of the previous MCDHF/RCI2 calculations. Comparing to experimental data, the accuracy of the MCDHF/RCI and MBPT calculations is at the same level. %The present two results are in better agreement with observations than those of ~\citet{Quinet.2011.V44.p195007}.
We also make comparisons between the three calculated results (MCDHF/RCI2, RCI, and MBPT) and the MCDHF/RCI results for levels listed in Table~\ref{table1}. Our MBPT excitation energies agree closely with those from MCDHF/RCI calculations with the mean relative energy difference of 0.11~\%, whereas the mean difference between the two earlier calculations and the present MCDHF/RCI results are 0.24~\% for MCDHF/RCI2, and 0.52~\% for RCI.

\begin{figure}[ht!]
\centering
\includegraphics[height=4.0in]{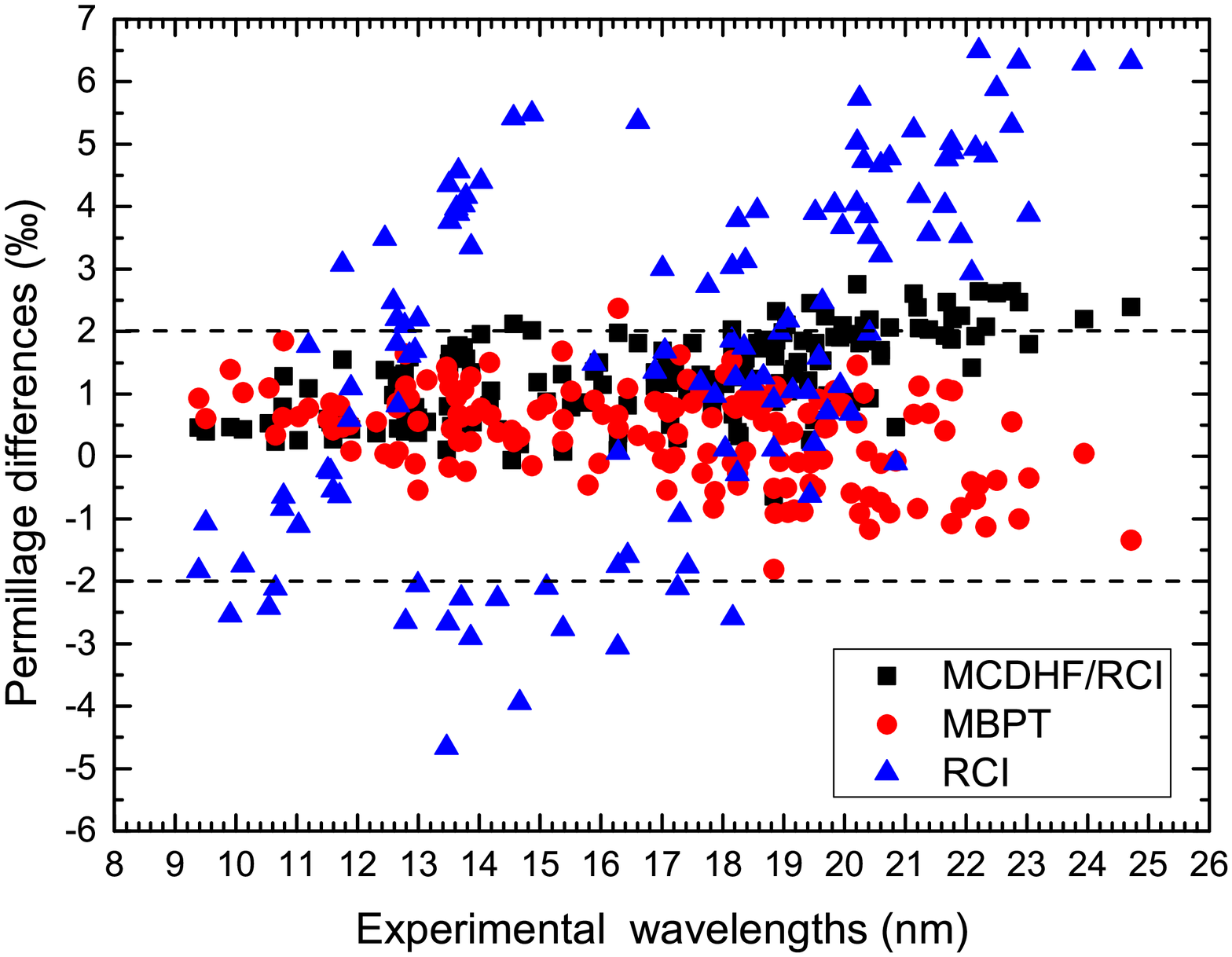}
\caption{Permillage differences between the experimental and theoretical wavelengths for the forbidden transitions of Hf ($Z=72$), Ta ($Z=73$), W ($Z=73$) and Au ($Z=79$) ions.  The squares, the circles, and the up-triangles stand for the permillage differences between the experimental wavelengths and the present MCDHF/RCI ones, the present MBPT ones, and the RCI ones \citep{Ralchenko.2011.V83.p32517,Osin.2012.V66.p286}, respectively. The experimental wavelengths are those from~Ralchenko $et~al.$~\cite{Ralchenko.2011.V83.p32517,Osin.2012.V66.p286}. The horizontal lines indicate the~0.2~\% differences of the experimental and theoretical wavelengths. }
\label{Figure01}
\end{figure}

To further gauge the quality of the MCDHF/RCI atomic data, in Table~\ref{table3} our MCDHF/RCI wavelengths of Hf ($Z=72$), Ta ($Z=73$), W ($Z=74$), and Au ($Z=79$) ions are compared with existing experimental data, as well as with the present MBPT values and the previous calculated results. These experimental values were obtained by~Ralchenko $et~al.$~\cite{Ralchenko.2011.V83.p32517,Osin.2012.V66.p286} who observed a total of 158 M1 transitions from extreme ultraviolet EBIT spectra between 9 nm and 25 nm. Our  MCDHF and MBPT results are in excellent agreement with observed values, i.e., better than 0.2~\% for most transitions, as illustrated in Figure~\ref{Figure01} showing the differences as a function of experimental values. For a total of 158 M1 transitions in Table~\ref{table3}, our MCDHF and MBPT results agree observed values better than 0.2~\% for 132 and 157 transitions, respectively.  
For these transitions,~Ralchenko $et~al.$~\cite{Ralchenko.2011.V83.p32517,Osin.2012.V66.p286} also reported wavelengths computed using the RCI method developed by ~\citet{Gu.2008.V86.p675}. Even if the general agreement between those values and experimental data is rather good, larger discrepancies (up to 0.7~\%) can be observed. Because of more complete consideration of electron correlation effects in the present MCDHF/RCI calculations via including large configuration sets, as shown in Section~\ref{sect:cal}, our MCDHF/RCI results are slightly better than those of ~Ralchenko $et~al.$~\cite{Ralchenko.2011.V83.p32517,Osin.2012.V66.p286}, and the accuracy of the MCDHF/RCI and MBPT calculations is at the same level.

% In most cases, the present MCDHF/RCI and MBPT results agree very well with the measured wavelengths, the mean di ¦ËMCDF/¦Ëexp being found equal to 1.001 ¡À 0.002, as illustrated in figure 1 showing this ratio as a function of ¦ËMCDF.s

%To further gauge the quality of the present MCDHF/RCI atomic data, our calculated wavelengths are compared with the available experimental values in Figure~\ref{Figure01}. These experimental values were obtained by~Ralchenko $et~al.$~\cite{Ralchenko.2011.V83.p32517,Osin.2012.V66.p286} who observed a total of 158 M1 transitions in Hf ($Z=72$), Ta ($Z=73$), W ($Z=74$), and Au ($Z=79$) ions from extreme ultraviolet EBIT spectra between 9 and 25 nm. The present MBPT results are also included in Figure~\ref{Figure01} for comparison. Our present MCDHF and MBPT results are in excellent agreement with observed values, i.e., better than 0.2\% for most levels. For these ions,~Ralchenko $et~al.$~\cite{Ralchenko.2011.V83.p32517,Osin.2012.V66.p286} also reported wavelengths computed using the RCI method developed by ~\citet{Gu.2008.V86.p675}. Even if the general agreement between those values and experimental data is rather good, larger discrepancies (up to 0.7\%) can be observed. Because of sufficient electron correlation effects considered in the present MCDHF/RCI calculations via including large configurations, as shown in Section~\ref{sect:cal}, our MCDHF/RCI results are slightly better than those of ~Ralchenko $et~al.$~\cite{Ralchenko.2011.V83.p32517,Osin.2012.V66.p286}

\subsection{Transition rates}
In Table~\ref{table4} we present our wavelengths ($\lambda$), Babushkin form of radiative rates ($A$), weighted oscillator strengths (\emph{gf}), and line strengths ($S$) from the MCDHF/RCI calculations, for 11988 M1 transitions among all levels of the $3s^2 3p^6 3d^k$ ($k=1-9$) ground configurations of highly charged ions ($Z=72-83$). The corresponding results for 18192 E2 transitions are given in Table~\ref{table5}. The indices adopted for the lower and upper levels of a transition are given in Table~\ref{table2}.

\begin{figure}[ht!]
\centering
\includegraphics[height=4.0in]{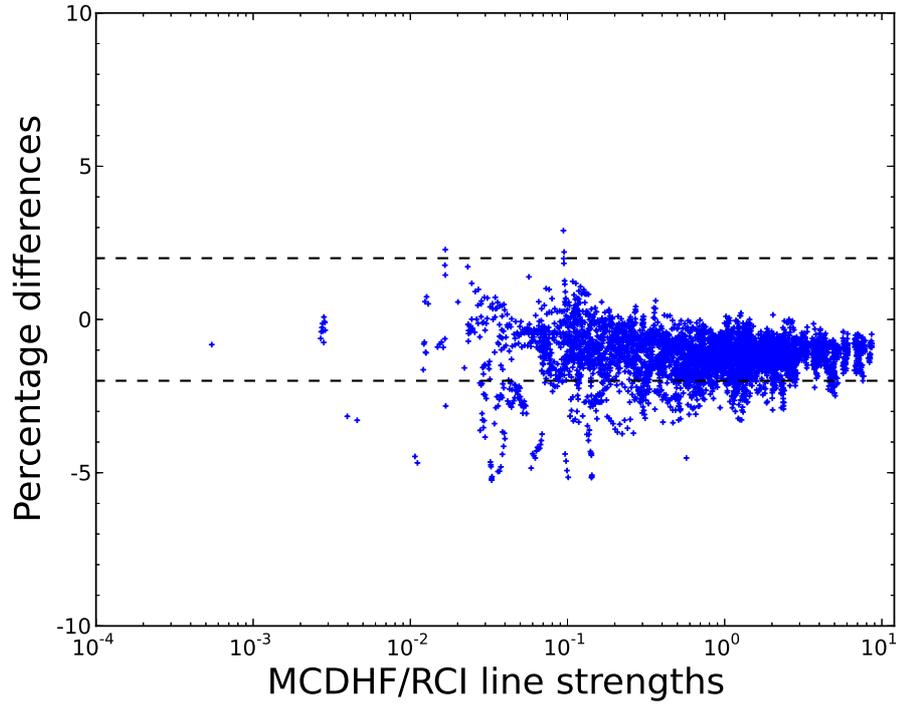}
\caption{Percentage differences of the M1 line strengths from the present MCDHF/RCI and MBPT calculations for 5004 relatively strong transitions. The horizontal line indicates the~2~\% differences of the two calculations.}
\label{Figure02}
\end{figure}

Since the M1 transitions are comparatively more important and more easily observed in experiment, we discuss these transitions in detail. Furthermore, the corresponding values from the MBPT calculations are not included in Tables~\ref{table4} and~\ref{table5} because the
results obtained are similar to those from the MCDHF/RCI calculations for a majority of transitions - for examples, see Table~\ref{table3} and Figure~\ref{Figure02}.

Figure~\ref{Figure02} compares the M1 line strengths from the present MCDHF/RCI and MBPT calculations for 5004 relatively strong transitions (\emph{A}-values $ \geq 10^{5}~s^{-1}$). The two sets of data agree within 2~\% for 4479 of 5004 transitions , within 5.2~\% for all transitions. This is highly satisfactory. However,
differences between the two calculations for weaker transitions ($< 10^{5}~s^{-1}$) are often larger and up to several orders of magnitude. The differences arise from different amount of electron correlation included in the calculations of the lower and upper levels of the transitions, and the weak transitions are usually more sensitive to cancellation among mixing coefficients. Therefore, their accuracy is always doubtful, as also discussed in detail by~\citet{Hibbert.2005.VT120.p71}. However, all of these transitions have very small \emph{A}-values and cannot be easily observed in experiments. Furthermore their contribution to level-population kinetics is much less important in comparison to those from the stronger transitions. Therefore, the larger differences between the different calculations for some weaker transitions do not affect the overall accuracy of the results. A similar analysis made for the E2 transitions shows that the MCDHF/RCI and MBPT calculations for all relatively strong  transitions agree within 5~\%. Therefore, the accuracy of the E2 transitions is comparable to those of the M1 ones discussed above.

%In Table~\ref{table3} transition rates for transitions between levels in the 2s22p2, 2s2p3, and 2p4 configurations are shown.
In Table~\ref{table3} the transition rates for Hf ($Z=72$), Ta ($Z=73$), W ($Z=74$) and Au ($Z=79$) ions from our present MCDHF/RCI calculations are compared with calculated rates from our MBPT calculations, as well as with the values from the RCI calculations reported by \citet{Osin.2012.V66.p286}. These transitions listed Table~\ref{table3} have been observed by~Ralchenko $et~al.$~\cite{Ralchenko.2011.V83.p32517,Osin.2012.V66.p286}. The agreement between the MCDHF/RCI and MBPT results is within 4~\% for all transitions listed in Table~\ref{table3}. Similar agreement can be found between the present and previous calculations, expect for the $(d_{+}^{2})_{4}-((d_{-}^{3})_{3/2},(d_{+}^{3})_{5/2})_{4}$ transition in the Cr-like Au ion. For this transition, the RCI value is an order of magnitude greater than those from the MCDHF/RCI and MBPT calculations, which is due to a misprint in Ref.~\citep{Osin.2012.V66.p286}.

%Also included in Table~\ref{table3} is the ratio R of the obtained transition probabilities in length and velocity gauge. For accurate wave functions and strong transitions the ratio R is expected to be close to unity, whereas ratios far from 1 are often associated with weaker transitions where uncertainties normally are larger. On the other hand, ratios close to unity do not necessarily imply accurate wave functions, but together with a thorough validation of calculated level energies these are still strong indicators on the quality of the calculations. As seen in Table~\ref{table3} the ratios are almost exclusively close to or at unity with only a few exceptions where the transition rates are smaller.
The uncertainties for line strengths in the last column of Tables~\ref{table4} and~\ref{table5} were calculated as follows. For each transition the absolute value of the difference of the MCDHF and MBPT $S$-values $\delta S_{ij}$ was determined. Then, the average of those $\delta S_{ij}$ values,  $\delta S_{av}$ was calculated for each decade of $S$-values (i.e., 1  $< S_{ij} < $ 10, 0.1  $< S_{ij} <$ 1, etc.). Finally, the larger of $\delta S_{ij}$ and $\delta S_{av}$ was accepted as the uncertainty of each particular line strength.

\section{Conclusions}\label{sect:sum}
In the present work, we carry out systematic calculations for excitation energies and radiative transition properties of the $3s^2 3p^6 3d^k$ ($k=1-9$) ground configurations of highly charged ions ($Z=72-83$). The calculations were performed by using the MCDHF and subsequent RCI approach (MCDHF/RCI) implemented within a new release~\cite{Jonsson.2013.V184.p2197} of the GRASP2K code~\cite{Jonsson.2007.V177.p597}, and a very large configuration expansion is used to take into account electron correlation effects.

The present MCDHF/RCI calculations provide a complete and consistent dataset for all levels of of the $3s^2 3p^6 3d^k$ ($k=1-9$) ground configurations of these highly charged ions. Comparisons with available observations and the present MBPT and previous calculations show that the present MCDHF/RCI results are highly accurate: for most excitation energies, the uncertainty should be less than 0.2~\%; for transition rates, the accuracy is better than 5~\% for most strong transitions. Since considerably larger amount of correlation effects are considered in the present MCDHF/RCI calculations, the resulting values represent an improvement in accuracy compared with earlier MCDHF/RCI works such as \citet{Quinet.2011.V44.p195007}. Comparing with the present MBPT calculations, the accuracy of the two calculations is at the same level. Thus the present work has significantly increased the amount of available accurate data for forbidden transitions within the $3s^2 3p^6 3d^k$ ($k=1-9$) ground configurations in highly charged ions.

\ack The authors acknowledge the support of the National Natural Science Foundation of China (Grant No. 11674066, No.~21503066, 11504421, and No.~11474034). This work is also supported by China Scholarship Council (Grant No. 201608130201), the Chinese Association of Atomic and Molecular Data,  Chinese National Fusion Project for ITER No.2015GB117000, and China Post- doctoral Science Foundation (Grant No. 2016M593019). The authors express their gratitude to Dr. Alexander Kramida of the National Institute of Standards and Technology, Gaithersburg, USA for the helpful discussions and detailed modifications on this paper. J. Yan would especially like to acknowledge the support from the visiting researcher program at the National Institute of Standards and Technology. K. Wang and S. Li express their gratitude for the support from the visiting researcher program at the Fudan University.

\clearpage
\bibliographystyle{adndt}
\bibliography{ref}
\clearpage

\section*{Figures}
\clearpage

\clearpage

\section*{Table ~\ref{table1}.\label{tbl1te} Energy levels of K-like W$^{55+}$ , Ti-like W$^{52+}$, and Fe-like W$^{48+}$ ions relative to the ground level (in cm$^{-1}$). }
\begin{tabular*}{0.95\textwidth}{@{}@{\extracolsep{\fill}}lp{5.5in}@{}}
Sequence        & Isoelectronic sequence                               \\
Key             &A number assigned to each level \\
Level           &The \emph{jj} designation of the level, with $\rm 1s^2 2s^2 2p^6 3s^2 3p^6$ omitted. In the notations adopted, $d_{-}$ corresponds to the principal quantum number $n=3$ and  $j_{-} = 3/2$, while $d_{+}$ is written for $n=3$ and $j_{+} = 5/2$. The subscript after the right parenthesis is the total $J$ value of the terms inside the parentheses. \\
DF & Dirac-Fock excitation energies \\
Electron correlation & Excitation energies obtained in each step of the MCDHF calculations, where $n = 3$ denotes the orbital set with the maximum principal quantum number $n = 3$, etc. \\
MCDHF/RCI & Excitation energies from the present MCDHF/RCI calculations\\
MBPT & Excitation energies from the present MBPT calculations\\
Exp. &  Experimental excitation energies from~Ralchenko $et~al.$~\cite{Ralchenko.2011.V83.p32517}\\
RCI & Excitation energies from the previous RCI calculations by~Ralchenko $et~al.$~\cite{Ralchenko.2011.V83.p32517}\\
MCDHF/RCI2 & Excitation energies from the previous MCDHF/RCI calculations  by ~\citet{Quinet.2011.V44.p195007}\\
\end{tabular*}
\clearpage

\section*{Table ~\ref{table2}.\label{tbl2te} Energy levels of the $3s^2 3p^6 3d^k$ ($k=1$-$9$) ground configurations of highly charged ions ($Z=72$-$83$) relative to the ground state (in cm$^{-1}$) from the present MCDHF/RCI calculations.}
\begin{tabular*}{0.95\textwidth}{@{}@{\extracolsep{\fill}}lp{5.5in}@{}}
Sequence        & Isoelectronic sequence \\
Z               & The nuclear charge      \\
Key             &A number assigned to each level \\
Level           &The \emph{jj} designation of the level, with $\rm 1s^2 2s^2 2p^6 3s^2 3p^6$ omitted. In the notations adopted, $d_{-}$ corresponds to the principal quantum number $n=3$ and  $j_{-} = 3/2$, while $d_{+}$ is written for $n=3$ and $j_{+} = 5/2$. The subscript after the right parenthesis is the total $J$ value of the terms inside the parentheses. \\
$J^\pi$ & The symmetry $J^{\pi}$ of each level, i.e.  the total angular momentum and parity\\
Energy(cm$^{-1}$) & Level energies relative to the ground state\\
\end{tabular*}

\clearpage
\section*{Table ~\ref{table3}.\label{tbl3te} Comparisons of wavelengths (nm) and transition rates ($\rm s^{-1}$) for some forbidden transitions of Hf ($Z=72$), Ta ($Z=73$), W ($Z=74$) and Au ($Z=79$)ions. These transitions have been observed by~Ralchenko $et~al.$~\cite{Ralchenko.2011.V83.p32517,Osin.2012.V66.p286}.}
\begin{tabular*}{0.95\textwidth}{@{}@{\extracolsep{\fill}}lp{5.5in}@{}}
Z               & The nuclear charge      \\
Sequence        & Isoelectronic sequence \\
i               & The number of the lower level \\
Lower level     & The \emph{jj} designation of the lower level \\
j               & The number of the upper level \\
Lower level     & The \emph{jj} designation of the upper level \\
Exp.            & The experimental results obtained by~Ralchenko $et~al.$~\cite{Ralchenko.2011.V83.p32517,Osin.2012.V66.p286}\\
MCDHF/RCI       & The results of the present MCDHF/RCI calculations\\
MBPT            & The results of the present MBPT calculations\\
RCI             & The results of the RCI calculations by~Ralchenko $et~al.$~\cite{Ralchenko.2011.V83.p32517,Osin.2012.V66.p286} \\
\end{tabular*}

\clearpage
\section*{Table ~\ref{table4}.\label{tbl4te} Wavelengths($\lambda$, in {nm}), transition rates(\emph{A}, in s$^{-1}$),  weighted oscillator strengths (\emph{gf}, dimensionless) and line strengths (\emph{S}, in atomic unit)  with the uncertainties (in \%), for 11988 M1 transitions among all levels of the $3s^2 3p^6 3d^k$ ($k=1$-$9$) ground configurations of highly charged ions ($Z=72$-$83$).}
\begin{tabular*}{0.95\textwidth}{@{}@{\extracolsep{\fill}}lp{5.5in}@{}}
Sequence        & Isoelectronic sequence \\
Z               & The nuclear charge      \\
i               & The number of the lower level \\
Lower level     & The \emph{jj} designation of the lower level \\
j               & The number of the upper level \\
Lower level     & The \emph{jj} designation of the upper level \\
$\lambda$       & Wavelengths of the present MCDHF/RCI calculations\\
\emph{A}        & Transition rates of the present MCDHF/RCI calculations\\
\emph{gf}       & Weighted oscillator strengths of the present MCDHF/RCI calculations\\
\emph{S}            & Line strengths of the present MCDHF/RCI calculations\\
Unc.             & Uncertainties  (in \%) of line strengths\\
\end{tabular*}

\clearpage
\section*{Table ~\ref{table5}.\label{tbl5te}Wavelengths ($\lambda$, in {nm}), transition rates (\emph{A}, in s$^{-1}$), weighted oscillator strengths (\emph{gf}, dimensionless) and line strengths (\emph{S}, in atomic unit)  with the uncertainties (in \%), for 18192 E2 transitions among all levels of the $3s^2 3p^6 3d^k$ ($k=1$-$9$) ground configurations of highly charged ions ($Z=72$-$83$).}
\begin{tabular*}{0.95\textwidth}{@{}@{\extracolsep{\fill}}lp{5.5in}@{}}
Sequence        & Isoelectronic sequence \\
Z               & The nuclear charge      \\
i               & The number of the lower level \\
Lower level     & The \emph{jj} designation of the lower level \\
j               & The number of the upper level \\
Lower level     & The \emph{jj} designation of the upper level \\
$\lambda$       & Wavelengths of the present MCDHF/RCI calculations\\
\emph{A}        & Transition rates of the present MCDHF/RCI calculations\\
\emph{gf}       & Weighted oscillator strengths of the present MCDHF/RCI calculations\\
\emph{S}            & Line strengths of the present MCDHF/RCI calculations\\
Unc.             & Uncertainties  (in \%) of line strengths\\
\end{tabular*}

\setlength{\tabcolsep}{0.5\tabcolsep}
\renewcommand{\arraystretch}{0.7}
\footnotesize % we need to squeeze the font size a lot!

\datatables % This command is necessary to get the table names in toc

%% If the table is to span over the whole text width, we set:
\setlength{\LTleft}{0pt} \setlength{\LTright}{0pt}

\setlength{\tabcolsep}{0.5\tabcolsep}
\renewcommand{\arraystretch}{1.0}
\scriptsize % we need to squeeze the font size a lot!

\begin{longtable}{@{\extracolsep\fill}ccrrrrrrrrrrrr@{}}
\caption{Energy levels of K-like W$^{55+}$ , Ti-like W$^{52+}$, and  Fe-like W$^{48+}$ ions relative to the ground state. See page\
\pageref{tbl1te} for Explanation of Tables\label{table1}}
    &      & & & \multicolumn{5}{c}{Electron correlation} & \multicolumn{2}{c}{Present calculations} & & \multicolumn{2}{c}{Other calculations}                                                              \\
\cline{5-9}\cline{10-11}\cline{13-14}\noalign{\vskip3pt}
Sequence&Key &Level & DF & $n=3$ &$n=4$& $n=5$& $n=6$& $n=7$& MCDHF/RCI & MBPT & Expt. &RCI &MCDHF/RCI2  \\
\hline\\
\endfirsthead\\
\caption[]{(\emph{continued})}
    &      & & & \multicolumn{5}{c}{Electron correlation} & \multicolumn{2}{c}{Present calculations} & & \multicolumn{2}{c}{Other calculations}                                                              \\
\cline{5-9}\cline{10-11}\cline{13-14}\noalign{\vskip3pt}
Sequence&Key &Level & DF & $n=3$ &$n=4$& $n=5$& $n=6$& $n=7$& MCDHF/RCI & MBPT & Expt. &RCI &MCDHF/RCI2  \\
\hline\\
\endhead
K  &    $ 1$ &   $(d_{-}^{1})_{3/2}                        $&$0      $&  $0      $&   $0      $&  $0      $&    $0      $&   $0         $ &    $0      $&  $0      $&     $      0   $&  $      0$&    $      0$\\
K  &    $ 2$ &   $(d_{+}^{1})_{5/2}                        $&$638774 $&  $639950 $&   $639569 $&  $639560 $&    $639577 $&   $639584    $ &    $625546 $&  $626558 $&     $ 626490   $&           &    $ 624700$\\
Ti &    $ 1$ &   $(d_{-}^{4})_{0}                          $&$0      $&  $0      $&   $0      $&  $0      $&    $0      $&                &    $0      $&  $0      $&     $       0.0$&           &    $      0$\\
Ti &    $ 2$ &   $((d_{-}^{3})_{3/2},(d_{+}^{1})_{5/2})_{1}$&$523541 $&  $530239 $&   $528965 $&  $529164 $&    $529273 $&                &    $516668 $&  $518082 $&     $  517630  $&           &    $ 516000$\\
Ti &    $ 3$ &   $((d_{-}^{3})_{3/2},(d_{+}^{1})_{5/2})_{4}$&$634757 $&  $643185 $&   $640196 $&  $639838 $&    $639845 $&                &    $613992 $&  $614779 $&                 &  $ 613000$&    $ 615600$\\
Ti &    $ 4$ &   $((d_{-}^{3})_{3/2},(d_{+}^{1})_{5/2})_{2}$&$655279 $&  $660378 $&   $657924 $&  $657430 $&    $657463 $&                &    $638623 $&  $639336 $&                 &  $ 638000$&    $ 639900$\\
Ti &    $ 5$ &   $((d_{-}^{3})_{3/2},(d_{+}^{1})_{5/2})_{3}$&$684343 $&  $689998 $&   $686803 $&  $686183 $&    $686185 $&                &    $666483 $&  $667034 $&     $  665562.1$&           &    $ 668600$\\
Ti &    $ 6$ &   $((d_{-}^{2})_{2  },(d_{+}^{2})_{2  })_{0}$&$1121586$&  $1127856$&   $1125942$&  $1126136$&    $1126261$&                &    $1103102$&  $1104654$&                 &  $1100000$&    $1101600$\\
\hline
\end{longtable}
\clearpage

\begin{longtable}{@{\extracolsep\fill}ccclrr@{}}
\caption{Energy levels from the present MCDHF/RCI calculations. See page\
\pageref{tbl2te} for Explanation of Tables\label{table2}}
Sequence&Z & Key &Level & $J^\pi$ & Energy(cm$^{-1}$)  \\
\hline\\
\endfirsthead\\
\caption[]{(\emph{continued})}
Sequence&Z & Key &Level & $J^\pi$ & Energy(cm$^{-1}$) \\
\hline\\
\endhead
$ K$&$ 72$&$   1$&$  (d_{-}^{1})_{3/2}                                          $&$   3/2^{ e}$&$        0.00$\\
$ K$&$ 72$&$   2$&$  (d_{+}^{1})_{5/2}                                          $&$   5/2^{ e}$&$   549756.52$\\
$ K$&$ 73$&$   1$&$  (d_{-}^{1})_{3/2}                                          $&$   3/2^{ e}$&$        0.00$\\
$ K$&$ 73$&$   2$&$  (d_{+}^{1})_{5/2}                                          $&$   5/2^{ e}$&$   586724.05$\\
$ K$&$ 74$&$   1$&$  (d_{-}^{1})_{3/2}                                          $&$   3/2^{ e}$&$        0.00$\\
$ K$&$ 74$&$   2$&$  (d_{+}^{1})_{5/2}                                          $&$   5/2^{ e}$&$   625546.08$\\
$ K$&$ 75$&$   1$&$  (d_{-}^{1})_{3/2}                                          $&$   3/2^{ e}$&$        0.00$\\
$ K$&$ 75$&$   2$&$  (d_{+}^{1})_{5/2}                                          $&$   5/2^{ e}$&$   666286.16$\\
\hline
\end{longtable}
\clearpage

\begin{longtable}{@{\extracolsep\fill}cccccccccccccc@{}}
\caption{Comparisons of wavelengths (nm) and transition rates ($\rm s^{-1}$) for some forbidden transitions of Hf ($Z=72$), Ta ($Z=73$), W ($Z=74$) and Au ($Z=79$) ions. These transitions have been observed by~Ralchenko $et~al.$~\cite{Ralchenko.2011.V83.p32517,Osin.2012.V66.p286}. See page\
\pageref{tbl3te} for Explanation of Tables\label{table3}}
  &           &  &             &   &           &  \multicolumn{4}{c}{Wavelengths (nm)} &\multicolumn{3}{c}{Transition rates ($\rm s^{-1}$)}                                                             \\
\cline{7-10}\cline{11-13}\noalign{\vskip3pt}
Z & Sequence  &i & Lower level & j &Upper Level& Exp.&   MCDHF/RCI & MBPT & RCI &MCDHF/RCI  & MBPT & RCI \\
\hline\\
\endfirsthead\\
\caption[]{(\emph{continued})}
  &           &  &             &   &           &  \multicolumn{4}{c}{Wavelengths (nm)} &\multicolumn{3}{c}{Transition rates ($\rm s^{-1}$)}                                                             \\
\cline{7-10}\cline{11-13}\noalign{\vskip3pt}
Z & Sequence  &i & Lower level & j &Upper Level& Exp.&   MCDHF/RCI & MBPT & RCI &MCDHF/RCI  & MBPT & RCI \\
\hline\\
\endhead
$72$ &  K & $  1 $ &  $(d_{-}^{1})_{3/2}                                 $  &  $ 2  $ & $(d_{+}^{1})_{5/2}                             $  &       $18.158$ &   $18.1899$  & $18.1561$ &  $18.2132$ &  $1.77{\rm E+}06$ &$1.73{\rm E+}06$& $1.76{\rm E+}06$\\
$72$ &  Ca& $  1 $ &  $(d_{-}^{2})_{2}                                   $  &  $ 3  $ & $((d_{-}^{1})_{3/2},(d_{+}^{1})_{5/2})_{3}     $  &       $19.525$ &   $19.5605$  & $19.5151$ &  $19.6012$ &  $2.49{\rm E+}06$ &$2.45{\rm E+}06$& $2.48{\rm E+}06$\\
$72$ &  Ca& $  1 $ &  $(d_{-}^{2})_{2}                                   $  &  $ 4  $ & $((d_{-}^{1})_{3/2},(d_{+}^{1})_{5/2})_{2}     $  &       $16.884$ &   $16.9068$  & $16.8988$ &  $16.9070$ &  $1.24{\rm E+}06$ &$1.21{\rm E+}06$& $1.23{\rm E+}06$\\
$72$ &  Ca& $  2 $ &  $(d_{-}^{2})_{0}                                   $  &  $ 6  $ & $((d_{-}^{1})_{3/2},(d_{+}^{1})_{5/2})_{1}     $  &       $22.158$ &   $22.2007$  & $22.1428$ &  $22.2673$ &  $1.14{\rm E+}06$ &$1.12{\rm E+}06$& $1.13{\rm E+}06$\\
$72$ &  Sc& $  1 $ &  $(d_{-}^{3})_{3/2}                                 $  &  $ 2  $ & $((d_{-}^{2})_{2  },(d_{+}^{1})_{5/2})_{5/2}   $  &       $21.762$ &   $21.8028$  & $21.7386$ &  $21.8713$ &  $2.28{\rm E+}06$ &$2.25{\rm E+}06$& $2.26{\rm E+}06$\\
$72$ &  Sc& $  1 $ &  $(d_{-}^{3})_{3/2}                                 $  &  $ 3  $ & $((d_{-}^{2})_{2  },(d_{+}^{1})_{5/2})_{3/2}   $  &       $19.639$ &   $19.6692$  & $19.6381$ &  $19.6876$ &  $1.84{\rm E+}06$ &$1.81{\rm E+}06$& $1.83{\rm E+}06$\\
$72$ &  Sc& $  1 $ &  $(d_{-}^{3})_{3/2}                                 $  &  $ 6  $ & $((d_{-}^{2})_{2  },(d_{+}^{1})_{5/2})_{1/2}   $  &       $18.15 $ &   $18.1869$  & $18.1778$ &  $18.1838$ &  $6.88{\rm E+}05$ &$6.76{\rm E+}05$& $6.84{\rm E+}05$\\
\hline
\end{longtable}
\clearpage

\begin{longtable}{@{\extracolsep\fill}ccccccccccc@{}}
	\caption{{\textnormal {Wavelengths ($\lambda$, in {nm}), transition rates (\emph{A}, in s$^{-1}$), weighted oscillator strengths (\emph{gf}, dimensionless) and line strengths (\emph{S}, in atomic unit) with the uncertainties (in \%), for 11988 M1 transitions among all levels of the $3s^2 3p^6 3d^k$ ($k=1$-$9$) ground configurations of highly charged ions ($Z=72$-$83$)}}. Table~\ref{table4} is published in its entirety in the machine-readable format. The values for the K-like sequence are shown here for guidance regarding its form and content. See page\
		\pageref{tbl4te} for Explanation of Tables\label{table4}}
	Sequence&Z&i &Lower level & j & Upper level &$\lambda$ &\emph{A}& \emph{gf}& \emph{S} & Unc.\\
	\hline\\
	\endfirsthead\\
	\caption[]{(\emph{continued})}
	Sequence&Z&i &Lower level & j & Upper level &$\lambda$ &\emph{A}& \emph{gf}& \emph{S} & Unc.\\
	\hline\\
	\endhead
	$ K$&$   72$&$   1    $&$(d_{-}^{1})_{3/2}                            $&$2    $&$(d_{+}^{1})_{5/2}                                      $&$1.8190{\rm E+}00$&$  1.77{\rm E+}06$&$  5.27{\rm E-}05$&$       2.37{\rm E+}00$& $1.4$\\
	$ K$&$   73$&$   1    $&$(d_{-}^{1})_{3/2}                            $&$2    $&$(d_{+}^{1})_{5/2}                                      $&$1.7044{\rm E+}00$&$  2.15{\rm E+}06$&$  5.62{\rm E-}05$&$       2.37{\rm E+}00$& $1.5$\\
	$ K$&$   74$&$   1    $&$(d_{-}^{1})_{3/2}                            $&$2    $&$(d_{+}^{1})_{5/2}                                      $&$1.5986{\rm E+}00$&$  2.61{\rm E+}06$&$  5.99{\rm E-}05$&$       2.37{\rm E+}00$& $1.4$\\
	$ K$&$   75$&$   1    $&$(d_{-}^{1})_{3/2}                            $&$2    $&$(d_{+}^{1})_{5/2}                                      $&$1.5009{\rm E+}00$&$  3.15{\rm E+}06$&$  6.38{\rm E-}05$&$       2.37{\rm E+}00$& $1.4$\\
	$ K$&$   76$&$   1    $&$(d_{-}^{1})_{3/2}                            $&$2    $&$(d_{+}^{1})_{5/2}                                      $&$1.4104{\rm E+}00$&$  3.79{\rm E+}06$&$  6.78{\rm E-}05$&$       2.37{\rm E+}00$& $1.4$\\
	$ K$&$   77$&$   1    $&$(d_{-}^{1})_{3/2}                            $&$2    $&$(d_{+}^{1})_{5/2}                                      $&$1.3266{\rm E+}00$&$  4.55{\rm E+}06$&$  7.21{\rm E-}05$&$       2.36{\rm E+}00$& $1.2$\\
	$ K$&$   78$&$   1    $&$(d_{-}^{1})_{3/2}                            $&$2    $&$(d_{+}^{1})_{5/2}                                      $&$1.2490{\rm E+}00$&$  5.45{\rm E+}06$&$  7.65{\rm E-}05$&$       2.36{\rm E+}00$& $1.2$\\
	$ K$&$   79$&$   1    $&$(d_{-}^{1})_{3/2}                            $&$2    $&$(d_{+}^{1})_{5/2}                                      $&$1.1768{\rm E+}00$&$  6.52{\rm E+}06$&$  8.12{\rm E-}05$&$       2.36{\rm E+}00$& $1.2$\\
\hline
\end{longtable}
\clearpage
\begin{longtable}{@{\extracolsep\fill}ccccccccccc@{}}
	\caption{{\textnormal { Wavelengths ($\lambda$, in {nm}), transition rates (\emph{A}, in s$^{-1}$), weighted oscillator strengths (\emph{gf}, dimensionless) and line strengths (\emph{S}, in atomic unit)  with the uncertainties (in \%), for 18192 E2 transitions among all levels of the $3s^2 3p^6 3d^k$ ($k=1$-$9$) ground configurations of highly charged ions ($Z=72$-$83$)}}. Table~\ref{table5} is published in its entirety in the machine-readable format. The values for the K-like sequence are shown here for guidance regarding its form and content. See page\
		\pageref{tbl5te} for Explanation of Tables\label{table5}. }
	Sequence&Z&i &Lower level & j & Upper level &$\lambda$ &\emph{A}& \emph{gf}& \emph{S}& Unc. \\
	\hline\\                                                                                   
	\endfirsthead\\                                                                            
	\caption[]{(\emph{continued})}                                                             
	Sequence&Z&i &Lower level & j & Upper level &$\lambda$ &\emph{A}& \emph{gf}& \emph{S}& Unc.  \\
	\hline\\
	\endhead
	K&$   72$&$   1    $&$(d_{-}^{1})_{3/2}                               $&$  2    $&$(d_{+}^{1})_{5/2}                                     $&$1.8190{\rm E+}01 $&$     3.20{\rm E+}02$&$	9.52{\rm E-}09$&$	3.41{\rm E-}04$&$1.7$\\
	K&$   73$&$   1    $&$(d_{-}^{1})_{3/2}                               $&$  2    $&$(d_{+}^{1})_{5/2}                                     $&$1.7044{\rm E+}01 $&$     4.14{\rm E+}02$&$	1.08{\rm E-}08$&$	3.19{\rm E-}04$&$1.5$\\
	K&$   74$&$   1    $&$(d_{-}^{1})_{3/2}                               $&$  2    $&$(d_{+}^{1})_{5/2}                                     $&$1.5986{\rm E+}01 $&$     5.35{\rm E+}02$&$	1.23{\rm E-}08$&$	2.99{\rm E-}04$&$1.5$\\
	K&$   75$&$   1    $&$(d_{-}^{1})_{3/2}                               $&$  2    $&$(d_{+}^{1})_{5/2}                                     $&$1.5009{\rm E+}01 $&$     6.88{\rm E+}02$&$	1.39{\rm E-}08$&$	2.81{\rm E-}04$&$1.5$\\
	K&$   76$&$   1    $&$(d_{-}^{1})_{3/2}                               $&$  2    $&$(d_{+}^{1})_{5/2}                                     $&$1.4104{\rm E+}01 $&$     8.81{\rm E+}02$&$	1.58{\rm E-}08$&$	2.63{\rm E-}04$&$1.4$\\
	K&$   77$&$   1    $&$(d_{-}^{1})_{3/2}                               $&$  2    $&$(d_{+}^{1})_{5/2}                                     $&$1.3266{\rm E+}01 $&$     1.12{\rm E+}03$&$	1.78{\rm E-}08$&$	2.47{\rm E-}04$&$1.3$\\
	K&$   78$&$   1    $&$(d_{-}^{1})_{3/2}                               $&$  2    $&$(d_{+}^{1})_{5/2}                                     $&$1.2490{\rm E+}01 $&$     1.43{\rm E+}03$&$	2.01{\rm E-}08$&$	2.33{\rm E-}04$&$1.4$\\
	K&$   79$&$   1    $&$(d_{-}^{1})_{3/2}                               $&$  2    $&$(d_{+}^{1})_{5/2}                                     $&$1.1768{\rm E+}01 $&$     1.81{\rm E+}03$&$	2.26{\rm E-}08$&$	2.19{\rm E-}04$&$1.1$\\
\hline
\end{longtable}
\clearpage
\end{document}